\documentclass[aps,pra,twocolumn,showpacs,amssymb]{revtex4}
\usepackage{graphicx}
\usepackage{bm}

\newcommand{\ints}{\int_0^{\hspace{.5ex}\infty}
{^{\hspace{-2.8ex}\textstyle '}}\hspace{2ex}}
\newcommand{\intss}{\int_0^{\hspace{1ex}\infty}
{^{\hspace{-3.5ex}\textstyle ''}}\hspace{2ex}}
\newcommand{\Ints}{\int_0^{\hspace{.5ex}\infty}
{^{^{^{\hspace{-2.8ex}\textstyle '}}\hspace{1.5ex}}}}

\newcommand{\sumprime}{\hspace{-.5ex}{^{\textstyle'}}}

\begin{document}

\title{
Resonant Energy Exchange between Atoms in Dispersing and
Absorbing Surroundings $\!\!$\footnote{Proceedings of
ICQO'2002, Raubichi, to appear in Optics and Spectroscopy}
}

\author{Ho Trung Dung}
\altaffiliation[Also at ]{Institute of Physics, National Center
for Natural Sciences and Technology, 1 Mac Dinh Chi Street,
District 1, Ho Chi Minh city, Vietnam}

\author{Ludwig Kn\"{o}ll}

\author{Dirk-Gunnar Welsch}

\affiliation{
Theoretisch-Physikalisches Institut,
Friedrich-Schiller-Universit\"{a}t Jena,
Max-Wien-Platz 1, 07743 Jena, Germany}

\date{June 21, 2002}

\begin{abstract}
Within the framework of quantization of the macroscopic electromagnetic
field, a master equation describing both the resonant dipole-dipole
interaction (RDDI) and the resonant atom-field interaction (RAFI) in the
presence of
dispersing and absorbing macroscopic bodies is derived, with the
relevant couplings being expressed in terms of the surroundings-assisted
Green tensor. It is shown that under certain conditions the RDDI
can be regarded as being governed by an effective Hamiltonian.
The theory, which applies to both weak and strong atom-field coupling,
is used to study the resonant energy exchange between two
(two-level) atoms sharing initially a single excitation.
In particular, it is
shown that in the regime of weak atom-field coupling there is a time window,
where the energy transfer follows a transfer-rate law of the type obtained by
ordinary second-order perturbation theory.
Finally, the spectrum of the light emitted during the energy transfer is
studied and the line splittings are discussed.
\end{abstract}


\maketitle


\section{Introduction}
\label{Intro}

It is well known that an initially excited molecule
can irreversibly transfer its
energy to unexcited species located nearby -- a fundamental process
involved in ultrafast photochemistry, excitation transport
in Langmuir-Blodgett films, determination of molecular structures,
etc.. For (acceptor) molecules that exhibit a sufficiently broad
(quasi-)continuum of vibronic states, the energy transfer may
be regarded as being irreversible, and the transfer rate
can be calculated by using standard second-order perturbation
theory (see, e.g., \cite{Ho02} and references therein).
If instead of molecules, the donor and acceptor species are
(two-level) atoms, the transfer can be reversed in principle,
leading to a  back-and-forth energy exchange between the atoms
\cite{Lehmberg70}. This process can be thoroughly described
by a nonperturbative treatment. When the interatomic interaction
is weak, however, the probability of the return
transfer can be so small that the energy exchange is essentially
again a one-way process. In addition to that, the atoms always feature
some natural linewidths, which can be regarded as continua
of states. An aim of the present article
is to establish the connection
between the intermolecular energy transfer and
the spontaneous-decay-assisted interatomic energy exchange.

The underlying mechanism of
intermolecular and interatomic energy transfer can be both
RDDI and RAFI \cite{Agarwal74}.
In free space and for transfer distances $R$ below the wavelength of
real-photon emission, the RDDI described in terms of an instantaneous
Coulomb interaction potential with the familiar
$R^{-3}$ distance law is typically considered as the leading mechanism.
For transfer distances exceeding the wavelength of real-photon emission,
the RDDI may be thought of as resulting from an extended
off-resonant atom-field interaction, and the effect of
the RAFI needs to be considered carefully. In particular,
in the multipolar-coupling quantum electrodynamics \cite{Craig84},
the near-zone and far-zone results simply arise as limits
of a unified theory based
on the full atom-field coupling (in electric-dipole approximation).

Similar to the single-atom spontaneous decay,
the interatomic energy exchange can be strongly
affected by the presence of material surroundings.
While RDDI and RAFI in free space have been more or
less well understood, there have been a number of open problems
when material surroundings are present.
In particular, it is not clear whether the near-zone
$R^{-3}$ law is still generally valid.
Another problem that needs
investigation is the regime of strong RAFI, i.e.,
when suitably chosen material surroundings form a high-$Q$
(micro-)resonator, which gives rise
to a strong atom-field coupling accompanied by quantum
Rabi oscillations. These and related problems
have recently been very actual in view of proposals
to entangle atoms (see, e.g., \cite{Sorensen02}) and to
implement quantum logic gates
(see, e.g., \cite{Barenco95,Lukin00}), using atoms in a cavity,
impurities in the condensed phase, or quantum dots
in a semiconductor.

For two atoms that interact with cavity-type modes,
a Hamiltonian for treating both RDDI and
RAFI was suggested in \cite{Kurizki96}. In the
Hamiltonian, an effective dipole-dipole interaction
energy, which may be thought of as resulting
from a perturbative treatment of the
off-resonant atom-field coupling, is added
to the exact on-resonant atom-field interaction energy.
Unfortunately, the relevant coupling strengths are introduced
as parameters rather than physically determined system quantities.
Moreover, the underlying mode expansion
concept makes the theory inapplicable for the realistic case
of absorbing and dispersing material surroundings. Inclusion
of material absorption and dispersion is especially important
in designing quantum computing systems where dissipation
and decoherence have to be strictly kept under control.

In what follows we give a more rigorous approach
to the problem, basing
on a recently developed quantization scheme
for the electromagnetic field in the presence of (linear) absorbing and
dispersing macroscopic bodies (see \cite{Knoll01} and references
therein). In this scheme, an operator noise polarization source
is introduced into the Maxwell equations, and the
electromagnetic field operators are
given by a source-quantity representation in terms of the
classical Green tensor and a continuous set of bosonic vector fields,
which play the role of the fundamental variables of the
composite system of the electromagnetic field plus the media.
The Green tensor, which can be calculated in a purely classical manner,
is essentially determined by the (Kramers-Kronig consistent) space- and
frequency-dependent complex permittivity, which characterizes the
relevant properties of the macroscopic bodies. The medium-assisted
electromagnetic field can then be coupled to additional
atomic systems, and the minimal-coupling Hamiltonian can be constructed,
from which the corresponding multipolar-coupling Hamiltonian
can be derived by means of a unitary transformation.

By using the multipolar-coupling Hamiltonian in electric-dipole
approximation, in Section \ref{densmatrix}
an equation of motion for the reduced density operator of
a system that consists of two-level atoms and that
part of the medium-assisted electromagnetic field that
resonantly interacts with the atoms is derived.
It is shown that when the (shifted) atomic transition frequencies
differ by amounts small compared to the scale of the
frequency variation
of the Green tensor, the system can be regarded as being governed by
a Hamiltonian, with the RDDI being described by a well-defined
effective interaction energy.
The problem of resonant energy exchange between two two-level atoms
is studied in Section \ref{sysdy}, with special emphasis on the rate
regime. The spectrum of the light emitted during
the energy exchange is considered in Section \ref{spectrum}.
Finally, some concluding remarks are given in
Section \ref{remarks}.


\section{Reduced density operator equation}
\label{densmatrix}

A system of $N$ two-level atoms [positions ${\bf r}_A$, transition
frequencies $\omega_A$, transition dipole moments ${\bf d}_A$
\mbox{($A$ $\!=$ $\!1,2,...,N$)}] interacting with the
electromagnetic field via electric-dipole transitions
in the presence of dispersing and absorbing bodies can be
described by the multipolar-coupling Hamiltonian \cite{Knoll01,Ho02}
\begin{eqnarray}
\label{e2}
\lefteqn{
   \hat{H} = \int \!{\rm d}^3{\bf r}
   \! \int_0^\infty \!\!{\rm d}\omega \,\hbar\omega
   \,\hat{\bf f}^\dagger({\bf r},\omega){}\hat{\bf f}({\bf r},\omega)
    + \sum_{A} {\textstyle\frac{1}{2}}\hbar\omega_A \hat{\sigma}_{Az}
}
\nonumber\\[1ex]&&\hspace{8ex}
    - \sum_{A} \int_0^\infty {\rm d}\omega \left[
    \hat{\bf d}_A
    \underline{\hat{\bf E}}({\bf r}_A,\omega)
    + {\rm H.c.}\right],
\end{eqnarray}
where
\begin{equation}
\label{e2.1}
        \hat{\bf d}_A = {\bf d}_A\hat{\sigma}_A
        + {\bf d}_A^\ast \hat{\sigma}_A^\dagger
\end{equation}
and
\begin{equation}
\label{e3}
     \underline{\hat{\bf E}}({\bf r},\omega)
      \!=\! i \sqrt{\frac{\hbar}{\pi\varepsilon_0}}
     \frac{\omega^2}{c^2}
   \!\!\int \!\!{\rm d}^3{\bf r}'
   \sqrt{\varepsilon_{\rm I}({\bf r}',\omega)}
   \,\bm{G}({\bf r},{\bf r}',\omega)
   {}\hat{\bf f}({\bf r}',\omega).
\end{equation}
Here, $\hat{\bf f}({\bf r},\omega)$ and
$\hat{\bf f}^\dagger({\bf r},\omega)$ are bosonic fields which
play the role of the fundamental variables of the electromagnetic
field and the medium, including a reservoir necessarily associated
with the losses in the medium, $\bm{G}({\bf r},{\bf r}',\omega)$
is the classical Green tensor, and
\mbox{$\varepsilon({\bf r},\omega)$
$\!=$ $\!\varepsilon_{\rm R}({\bf r},\omega)$
$\!+$ $\!i\varepsilon_{\rm I}({\bf r},\omega)$} is the complex
permittivity. There are no direct
Coulomb forces between the particles in the multipolar-coupling
Hamiltonian (\ref{e2}); all interactions are mediated by
the medium-assisted electromagnetic field.

With regard to the interaction of the atoms with the
electromagnetic field, it is convenient to decompose the latter
into an on-resonant part (denoted by $\ints {\rm d\omega}\ldots$)
and an off-resonant part (denoted by$\intss {\rm d\omega}\ldots$).
Let us consider the temporal evolution of the system that consists
of the atoms and the on-resonant part of the electromagnetic field.
For this purpose we formally solve the Heisenberg equation of motion
for $\hat{\bf f}({\bf r},\omega)$
[and $\hat{\bf f}^\dagger({\bf r},\omega)$] and
insert the result into the equation of motion for
an arbitrary system operator.
Applying a (coarse-grained)
Markov approximation to the slowly varying atomic variables
in the off-resonant frequency integrals,
and assuming that the off-resonant free-field is initially
\mbox{($t$ $\!=$ $\!0$)}
prepared in the vacuum state, after some algebra,
we derive the following equation of motion for
the system density operator in the Schr\"odinger picture:
\begin{eqnarray}
\label{e10.2d}
\lefteqn{
       \,\dot{\!\hat{\varrho}} =
       -\frac{i}{\hbar} \left[\,\hat{\!\tilde{H}}_{\rm S},
       \hat{\varrho}\right]
}
\nonumber\\[1ex]&&\hspace{4ex}
       +\, \Biggl\{ i\sum_{A,A'}\sumprime
       \left[
       \delta^-_{A^\ast A'}
          \left(\hat{\sigma}^\dagger_A\hat{\sigma}_{A'}\hat{\varrho}
	  - \hat{\sigma}_{A'}\hat{\varrho}\hat{\sigma}^\dagger_A\right)
       \right.
\nonumber\\[1ex]&&\hspace{8ex}
        \left.
        +\,\delta^+_{A A^{'\ast}}
	   \left(\hat{\sigma}_A\hat{\sigma}^\dagger_{A'}\hat{\varrho}
	   -\hat{\sigma}^\dagger_{A'}\hat{\varrho}\hat{\sigma}_A\right)
        \right]
	+ {\rm H.c.}
	\Biggr\},
\end{eqnarray}
where
\begin{eqnarray}
\label{e3.1a}
\lefteqn{
   \hat{\tilde{H}}_{\rm S} = \int \!{\rm d}^3{\bf r}
   \! \Ints \!\!{\rm d}\omega \,\hbar\omega
   \,\hat{\bf f}^\dagger({\bf r},\omega){}\hat{\bf f}({\bf r},\omega)
    + \sum_{A} {\textstyle\frac{1}{2}}
      \hbar\tilde{\omega}_A \hat{\sigma}_{Az}
}
\nonumber\\[1ex]&&\hspace{10ex}
    - \sum_{A} \Ints {\rm d}\omega \left[
    \hat{\bf d}_A
    \underline{\hat{\bf E}}({\bf r}_A,\omega)
    + {\rm H.c.}\right],
\end{eqnarray}
and the notation $\sum_{A,A'}'$ indicates that \mbox{$A$ $\!\neq$ $A'$}.
In Eqs. (\ref{e10.2d}) and (\ref{e3.1a}),
the shifted atomic transition frequencies
and the resonant dipole-dipole coupling strengths
are defined according to
\begin{equation}
\label{e10}
    \tilde{\omega}_A = \omega_A-\delta_{A^\ast A}.
\end{equation}
\begin{equation}
\label{e3.8}
       \delta_{A^\ast A} = \delta^-_{A^\ast A} -  \delta^+_{A^\ast A} ,
\end{equation}
\begin{eqnarray}
\label{e3.7}
       \delta^{-(+)}_{AA'}
       = \frac{\cal P}{\pi\hbar\varepsilon_0}
       \int_0^\infty {\rm d}\omega \,\frac{\omega^2}{c^2}
       \frac{ {\bf d}_A \,{\rm Im}\,\bm{G} ({\bf r}_A,{\bf r}_{A'},\omega)
       \,{\bf d}_{A'} }
       {\omega-(+)\tilde{\omega}_{A'}}
\end{eqnarray}
[${\cal P}$ -- principal value].
The notation $A^\ast$ ($A'^\ast$) means that ${\bf d}_A$
(${\bf d}_{A'}$) in
Eq. (\ref{e3.7}) has to be replaced with
its complex conjugate ${\bf d}_A^\ast$ (${\bf d}_{A'}^\ast$).
In Eq. (\ref{e10.2d}), both the RDDI
and the RAFI are taken into account,
without any restriction to the strength of the latter one.
Applying the Kramers-Kronig relation to the Green tensor,
we find from Eq.~(\ref{e3.7}) that the relations
\begin{equation}
\label{e10.1}
       \delta_{A^\ast A} =
       \frac{\omega_A^2}{\hbar\varepsilon_0 c^2}\,
       {\bf d}_A^\ast {\rm Re}\,\bm{G}({\bf r}_A,{\bf r}_A,
       \tilde{\omega}_{A}){\bf d}_A
       -  2\delta^+_{A^\ast A}
\end{equation}
and
\begin{eqnarray}
\label{e10.2}
\lefteqn{
       \left.\delta_{A^\ast A'}\right|_{A\neq A'}
       = \delta^-_{A^\ast A'} +  \delta^+_{A^\ast A'}
}
\nonumber\\[1ex]&&\hspace{2ex}
       = \frac{\tilde{\omega}_{A'}^2} {\hbar\varepsilon_0 c^2}\,
       {\bf d}_A^\ast {\rm Re}\,\bm{G}({\bf r}_A,{\bf r}_{A'}
       \tilde{\omega}_{A'}){\bf d}_{A'}
\end{eqnarray}
are (approximately) valid.

{F}rom Eq.~(\ref{e10.2}) it is seen
that when the transition frequencies $\tilde{\omega}_{A}$ and
$\tilde{\omega}_{A'}$ of two atoms $A$ and $A'$ are different
from each other, then
$|\delta_{A^\ast A'}|$ and $|\delta_{A'^\ast A}|$ are not
symmetric with respect to $A$ and $A'$. Only if the differences
\mbox{$|\tilde{\omega}_A$ $\!-$ $\!\tilde{\omega}_{A'}|$} are small
compared with the frequency scale of variation of the Green tensor,
so that
\begin{equation}
\label{e10.3}
\delta_{A^\ast A'}^\pm \simeq \delta_{A' A^\ast}^\pm
\end{equation}
holds, this asymmetry can be disregarded.
Then Eq.~(\ref{e10.2d}) reduces to
\begin{equation}
\label{e10.3b}
       \,\dot{\!\hat{\varrho}} =
       -\frac{i}{\hbar}
       \Bigl[\Bigl(
       \,\hat{\!\tilde{H}}_{\rm S}
       - \sum_{A,A'}\sumprime
       \hbar\delta_{A^\ast A'}
       \hat{\sigma}^\dagger_A\hat{\sigma}_{A'}
       \Bigr)
       ,\hat{\varrho}\Bigr].
\end{equation}
Recalling the definition of $\,\hat{\!\tilde{H}}_S$,
we see that the motion of the system is governed
by the effective Hamiltonian
\begin{eqnarray}
\label{e11}
\lefteqn{
   \hat{H}_{\rm eff} = \int \!{\rm d}^3{\bf r}
   \Ints {\rm d}\omega \,\hbar\omega
   \,\hat{\bf f}^\dagger({\bf r},\omega){}\hat{\bf f}({\bf r},\omega)
}
\nonumber\\[1ex]&&\hspace{7ex}
    + \sum_{A} {\textstyle\frac{1}{2}}\hbar\tilde{\omega}_A \hat{\sigma}_{Az}
    - \sum_{A,A'}\sumprime 
    \hbar \delta_{A^\ast A'}
       \hat{\sigma}^\dagger_A\hat{\sigma}_{A'}
\nonumber\\[1ex]&&\hspace{7ex}
    - \sum_{A} \Ints {\rm d}\omega \left[
    \hat{\bf d}_A  \underline{\hat{\bf E}}({\bf r}_A,\omega)
    + {\rm H.c.}\right],
\end{eqnarray}
with $\underline{\hat{\bf E}}({\bf r}_A,\omega)$ being given
by Eq.~(\ref{e3}).
The Hamiltonian (\ref{e11}) is remarkable in several respects. Firstly,
it applies to atoms surrounded by arbitrarily configured,
dispersive and absorptive media. Secondly, it goes beyond the rotating-wave
approximation. Thirdly, it contains a RDDI
energy that is explicitly expressed in terms of the
medium-assisted Green tensor, according to Eq.~(\ref{e10.2}).
Note that both the single-atom
transition frequency shift and the interatom RDDI
result from the off-resonant atom-field coupling.

Strong RDDI may be expected
if the atoms are sufficiently close to each other.
To give a simple example of the effect of material absorption, let us
assume that the atoms are embedded in bulk material of complex
permittivity $\varepsilon(\omega)$. Using the bulk-material
Green tensor (see, e.g., \cite{Knoll01}), from Eq.~(\ref{e10.2})
we find in the short-distance limit
\begin{equation}
\label{e12}
      \delta_{A^\ast A'} \!=\! \frac{1}{4\pi\hbar\varepsilon_0 R^3}
      {\rm Re}\!\left[\frac{1}{\varepsilon(\tilde{\omega}_{A'})}\right]\!\!
      \left( 3 \frac{{\bf d}_A^\ast {} {\bf R}}{R}\,
             \frac{{\bf d}_{A'} {} {\bf R}}{R}
             - {\bf d}_A^\ast {} {\bf d}_{A'}
      \!\right)
\end{equation}
(${\bf R}$ $\!=$ $\!{\bf r}_A$ $\!-$ $\!{\bf r}_{A'}$).
In free space, Eq.~(\ref{e12}) reduces to the well-known result
that the RDDI simply corresponds to the
(near-field) Coulomb-type interaction.
Note that though the characteristic $R^{-3}$ distance dependence
observed in free space is not changed by the bulk medium, it
may be changed by surroundings with more complex geometry.
In the long-distance limit, we find that
\begin{eqnarray}
\label{e12.1}
\lefteqn{
      \delta_{A^\ast A'} = \frac{c^2}{
      4\pi\hbar\varepsilon_0 R \tilde{\omega}_{A'}^2}
      \left( {\bf d}_A^\ast {} {\bf d}_{A'}
      - \frac{{\bf d}_A^\ast {} {\bf R}}{R}\,
             \frac{{\bf d}_{A'} {} {\bf R}}{R}
      \right)
}
\nonumber\\[1ex]&&\hspace{1ex}\times\,
      \cos\!\left[n_{\rm R}(\tilde{\omega}_{A'})
      \frac{\tilde{\omega}_{A'}R}{c}\right]
      \exp\!\left[- n_{\rm I}(\tilde{\omega}_{A'})
      \frac{\tilde{\omega}_{A'}R}{c}\right]
\end{eqnarray}
[\mbox{$n(\omega)$ $\!=$
$\!n_{\rm R}(\omega)$ $\!+$ $\!in_{\rm I}(\omega)$
$\!=$ $\!\sqrt{\varepsilon(\omega)}]$}, i.e.,
the (harmonically modulated)
$R^{-1}$ dependence observed in free space
is changed to an exponential decrease according to
$\exp[-n_{\rm I}(\tilde{\omega}_{A'})\tilde{\omega}_{A'}R/c]$
due to material absorption. Thus, material absorption
can drastically reduce the RDDI
strength with increasing mutual distance of the atoms.


\section{Resonant energy exchange by a two-atom system}
\label{sysdy}

Let us assume that initially the atoms share a single excitation
while the field is in the vacuum state.
In the Schr\"{o}dinger picture we may write, on omitting
off-resonant terms, the state vector of the system in the form of
\begin{eqnarray}
\label{e4.1}
\lefteqn{
    |\psi(t)\rangle = \sum_A C_A(t)
    e^{-i(\tilde{\omega}_A-\bar{\omega})t}
    |U_A\rangle |\{0\}\rangle
}
\nonumber\\[1ex]&&\hspace{-8ex}
     +\! \int \!{\rm d}^3{\bf r} \Ints \!\!{\rm d}\omega\,
     C_{Li}({\bf r},\omega,t)
     e^{-i (\omega-\bar{\omega})t}
     |L\rangle\hat{f}_i^\dagger({\bf r},\omega)|\{0\}\rangle
\end{eqnarray}
($\bar{\omega}$ $\!=$ $\!$ $\!\frac{1}{2}\sum_A\tilde{\omega}_A$).
Here, $|U_A\rangle$ is the atomic state with the $A$th atom
in the upper state and all the other atoms in the lower state,
and $|L\rangle$ is the atomic state with all atoms in the lower state.
Accordingly, $|\{0\}\rangle$ is the vacuum state of the
rest of the system, and
$\hat{f}^\dagger_i({\bf r},\omega) |\{0\}\rangle$
is a state, where a single quantum is excited.

{F}rom the effective Hamiltonian given in Eq.~(\ref{e11}),
we obtain the following
system of coupled integrodifferential equations for the
(slowly-varying) upper-state probability amplitudes $C_A$:
\begin{eqnarray}
\label{e9}
\lefteqn{
        \dot{C}_A(t) =
	  \sum_{\scriptstyle A' \atop \scriptstyle A'\neq A}
	i\delta_{A^\ast A'}
        e^{i(\tilde{\omega}_A-\tilde{\omega}_{A'})t}
        C_{A'}(t)
}
\nonumber \\[1ex]&&\hspace{0ex}
        + \sum_{A'}
        \int_0^t {\rm d}t'
        \Ints \!{\rm d}\omega
         K_{A^\ast A'}(t,t';\omega)
        \, C_{A'}(t'),
\end{eqnarray}
where
\begin{eqnarray}
\label{e6}
\lefteqn{
        K_{AA'}(t,t';\omega)
        = -\frac{1} {\hbar\pi\varepsilon_0}
        \biggl[ \frac{\omega^2}{ c^2}
        e^{-i(\omega-\tilde{\omega}_A)t}
        e^{i(\omega-\tilde{\omega}_{A'})t'}
}
\nonumber \\[1ex]&&\hspace{14ex}\times
       {\bf d}_A {\rm Im}\,\bm{G}({\bf r}_A,{\bf r}_{A'},\omega)
       {\bf d}_{A'} \biggr].
\end{eqnarray}
We see that the RAFI is determined by the imaginary part of the
Green tensor. In order to get
insight into the atomic motion on the basis of closed solutions,
let us consider two atoms and restrict our attention to
the limiting cases of weak and strong RAFI.


\subsection{Weak atom-field coupling}
\label{weakcoupling}

In the weak coupling regime, the integral expression in
Eqs.~(\ref{e9}) can be treated in a (coarse-grained) Markov
approximation.
We now make the simplifying assumption that the transition frequencies
of the two atoms are nearly equal to each other,
\mbox{$\tilde{\omega}_A$ $\!\simeq$ $\!\tilde{\omega}_B$}, but allow for
\mbox{$\Gamma_{A^\ast A}$ $\!\neq$ $\!\Gamma_{B^\ast B}$}.
The latter may happen, e.g., when the atoms
$A$ and $B$ have different dipole matrix elements and/or
different dipole orientations. It is then not difficult to
solve Eqs.~(\ref{e9}).
In particular, if the atom $B$ is initially in the lower state,
\mbox{$C_B(t$ $\!=$ $\!0)$ $\!=$ $\!0$}, we obtain
\begin{eqnarray}
\label{e16.1}
\lefteqn{
        C_A = \frac{1}{2D}\left\{
        \left[ - {\textstyle\frac{1}{2}}\left(
        \Gamma_{A^\ast A}-\Gamma_{B^\ast B}\right) + D \right]
        e^{D_+t/2}
        \right.
}
\nonumber\\[1ex]&&\hspace{8ex}
        \left.
        + \left[{\textstyle\frac{1}{2}}\left(
        \Gamma_{A^\ast A}-\Gamma_{B^\ast B}\right) + D \right] e^{D_-t/2}
        \right\},
\\[1ex]&&\hspace{0ex}
\label{e16.2}
        C_B = \frac{{\cal K}_{B^\ast A}}{D}
        \left(  e^{D_+t/2} - e^{D_-t/2}  \right),
\end{eqnarray}
where
\begin{eqnarray}
\label{e16}
&\displaystyle
\Gamma_{AA'} = \frac{2\tilde{\omega}_{A'}^2}{ \hbar\varepsilon_0c^2}\,
     {\bf d}_A \,{\rm Im}\,\bm{G}({\bf r}_A,{\bf r}_{A'},
     \tilde{\omega}_{A'})\, {\bf d}_{A'}\,,
\\[1ex]
\label{e15}
&\displaystyle
       {\cal K}_{AB}
       = - {\textstyle\frac{1}{2}} \Gamma_{AB} + i\delta_{AB}\,,
\\[1ex]
\label{e16.3}
&\displaystyle
        D = \left[{\textstyle\frac{1}{4}}
        (\Gamma_{A^\ast A}-\Gamma_{B^\ast B})^2
        +4{\cal K}_{A^\ast B}{\cal K}_{B^\ast A}\right]^{1/2} ,
\\[1ex]
\label{e16.4}
&\displaystyle
         D_\pm = -{\textstyle\frac{1}{2}}
         (\Gamma_{A^\ast A}+\Gamma_{B^\ast B}) \pm D .
\end{eqnarray}

When the two atoms are identical and have equivalent positions
and dipole orientations with respect to the material
surroundings
such that the relations
\begin{eqnarray}
\label{e16.4a}
&\displaystyle
        \Gamma_{A^\ast A}=\Gamma_{B^\ast B},
\\
\label{e16.4b}
&\displaystyle
        \Gamma_{A^\ast B}=\Gamma_{B^\ast A},\quad
	\delta_{A^\ast B}=\delta_{B^\ast A}
\end{eqnarray}
are valid, we have \mbox{${\cal K}_{A^\ast B}$ $\!=$ $\!{\cal K}_{B^\ast A}$},
and $\Gamma_{A^\ast B}$, $\Gamma_{B^\ast A}$,
$\delta_{A^\ast B}$, and $\delta_{B^\ast A}$ are real quantities
due to the reciprocity of the Green tensor.
Then Eqs.~(\ref{e16.1}) and (\ref{e16.2}) yield
\begin{equation}
\label{e16.5}
         P_{A(B)}(t)= {\textstyle \frac{1}{2}}
         \left[ \cosh\left(\Gamma_{A^\ast B}t\right)
       +(-) \cos\left(2\delta_{A^\ast B}t\right) \right]
       e^{-\Gamma_{B^\ast B} t}.
\end{equation}
[\mbox{$P_{A(B)}(t)$ $\!=|C_{A(B)}(t)|^2$}].
Eqs.~(\ref{e16.1}) -- (\ref{e16.4})
[and thus Eq.~(\ref{e16.5})] are valid for arbitrary dispersing
and absorbing material surroundings of the atoms. In particular,
substituting in Eqs.~(\ref{e16}) and (\ref{e15}) for the Green tensor
the vacuum Green tensor, Eq.~(\ref{e16.5}) reduces to that one obtained
in Ref.~\cite{Lehmberg70}. Accordingly, using the Green tensor
for absorbing bulk material, the result in Ref.~\cite{Juzeliunas94}
is recognized.

{F}rom Eq.~(\ref{e16.5}) a damped oscillatory excitation
exchange between the two atoms is seen.
For sufficiently small times, \mbox{$\Gamma_{B^\ast B}t$
$\!\ll$ $\!1$}, and strong RDDI,
\mbox{$|\delta_{A^\ast B}|\!$ $\!\gg \Gamma_{B^\ast B}$},
the oscillatory behavior
dominates. This can typically be observed when the atoms are
sufficiently near to each other, but can also be realized
for more moderate distances with the interatom coupling being
mediated by a high-$Q$ medium-assisted field resonance
\cite{Ho01a}. In the opposite limit of weak resonant dipole-dipole
coupling, $P_A(t)$ decreases monotonously while $P_B(t)$ features
one peak which separates the regime of energy transfer from atom
$A$ to atom $B$ at early times and the subsequent decay of the
excited state of atom $B$.


\subsection{Rate regime}
\label{rateregime}

As already mentioned, for weak RDDI
the energy transfer is one-way; that is, from atom $A$
to atom $B$. In this case, a transfer rate $w_1$ can be
defined according to
\begin{eqnarray}
\label{e16.5a}
      w_1=\frac{ {\rm d}P_B(t) }{ {\rm d}t} \biggr|_{t_0},
\end{eqnarray}
where $t_0$ is determined from the conditions that
\begin{eqnarray}
\label{e16.5b}
      \frac{ {\rm d}^2P_B(t) }{ {\rm d}^2t} \biggr|_{t_0} = 0,
\qquad
      \frac{ {\rm d}P_B(t) }{ {\rm d}t} \biggr|_{t_0} >0.
\end{eqnarray}

From Eq.~(\ref{e16.2}) [together with Eqs.~(\ref{e16.3}) and
(\ref{e16.4}) for negligibly small $K_{A^\ast B}$ and
$K_{B^\ast A}$ therein] it then follows that
\begin{eqnarray}
\label{e16.5c}
\lefteqn{
      w_1 \simeq \frac{|K_{B^\ast A}|^2}{ D^2} e^{D_-t_0}
}
\nonumber\\[1ex]&&\hspace{6ex}\times
      \left[ D_- + D_+ e^{2Dt_0}
      - (D_+ + D_-) e^{Dt_0} \right],
\end{eqnarray}
\begin{eqnarray}
\label{e16.5d}
\lefteqn{
      t_0 \simeq \frac{1}{ D}
      \ln\biggl( \frac{1}{ 4D_+^2} \Bigl\{(D_+ + D_-)^2
}
\nonumber\\[1ex]&&\hspace{6ex}
      - \, 2D\left[(D_++D_-)^2 +4D_+D_-\right]^{1/2} \Bigr\}
     \biggr).
\end{eqnarray}
Let us analyze Eq.~(\ref{e16.5c}) for three particular cases.
\begin{itemize}
\item[(i)]
If \mbox{$\Gamma_{A^\ast A}$ $\!\gg$ $\!\Gamma_{B^\ast B}$} is valid,
then from Eq.~(\ref{e16.5d}) it follows that
\mbox{$\Gamma_{A^\ast A}t_0$ $\!=$ $\!\ln 4$},
and Eq.~(\ref{e16.5c}) reduces to
\begin{equation}
\label{e16.7}
    w_1 = |{\cal K}_{B^\ast A}|^2/\Gamma_{A^\ast A}\,.
\end{equation}
\item[(ii)]
For \mbox{$\Gamma_{A^\ast A}$ $\!=$ $\!\Gamma_{B^\ast B}$}
the relation \mbox{$\Gamma_{B^\ast B}t_0$
$\!=$ $\!2$ $\!-$ $\!\sqrt{2}$} holds. Thus, Eq.~(\ref{e16.5c})
reads as
\begin{equation}
\label{e16.6}
    w_1 \simeq |{\cal K}_{B^\ast A}|^2 2\left(\sqrt{2}-1\right)
    e^{-\left(2-\sqrt{2}\right)}/\Gamma_{B^\ast B}\,.
\end{equation}
\item[(iii)]
In the case where \mbox{$\Gamma_{B^\ast B}$
$\!\gg$ $\!\Gamma_{A^\ast A}$} is valid, one finds
\mbox{$\Gamma_{B^\ast B}t_0$ $\!=$ $\!\ln 4$}, so that
Eq.~(\ref{e16.5c}) reduces to
\begin{equation}
\label{e16.7a}
        w_1=|{\cal K}_{B^\ast A}|^2/\Gamma_{B^\ast B}\,.
\end{equation}
\end{itemize}

For molecules, a rate regime is commonly
considered, and the
transfer rate is calculated by means of Fermi's golden rule
in second-order perturbation theory with regard to the
molecule-field interaction,
\begin{eqnarray}
\label{e25.2a}
      w=\sum_{f,i}p_iw_{fi},
\end{eqnarray}
where $w_{fi}$ is proportional to the absolute square of the
second-order interaction matrix element and the energy conserving
$\delta$-function \mbox{$\delta(\omega_f$ $\!-$ $\!\omega_i)$}, and the
sum runs over the (quasi-)continuum of initial ($i$) and/or
final ($f$) vibronic states.
In particular, short-distance energy transfer, where the intermolecular
coupling is essentially static, has been well known as F\"{o}rster
transfer \cite{Forster48,Dexter53}. Later on energy transfer over
arbitrary distances has been considered (see, e.g.,
\cite{Juzeliunas94,Ho02} and references therein).

For two two-level atoms, the single-transition probability per unit time
$w_{if}$ corresponds to $\dot{P}_{B}$. Starting from the effective
Hamiltonian (\ref{e11}) and evaluating $C_{B}(t)$ from
Eq.~(\ref{e9}) in first-order perturbation theory, we derive,
on making the standard long-time assumption,
\begin{equation}
\label{e28}
       \dot{P}_B = 2\pi \left|{\cal K}_{B^\ast A}\right|^2
       \delta\!\left(\tilde{\omega}_A-\tilde{\omega}_B\right),
\end{equation}
which is fully consistent with the result,
obtained in second-order perturbation theory on the basis of the
original (fundamental) Hamiltonian (\ref{e2}) \cite{Ho02}.
The only new feature in the current treatment is that
the medium induced atomic frequency shifts are
taken into account.

For establishing a rate regime, it is necessary that, according
to Eq.~(\ref{e25.2a}), a continuum of initial and/or final states
is involved in the transition. In the two-atom problem at hand,
the continua of states are obviously provided by the atomic level
broadening due to the spontaneous decay.
Assuming Lorentzian line shapes,
we may perform the integrals in Eq.~(\ref{e25.2a})
to obtain
\begin{equation}
\label{e28.4}
      w = \frac{4|{\cal K}_{B^\ast A}|^2 }{  \Gamma_{A^\ast A}
      +\Gamma_{B^\ast B}} \, p_A(\tilde{\omega}_A).
\end{equation}
Let us identify $p_A(\tilde{\omega}_{A})$ with $P_A(t_0)$
[Eq.~(\ref{e16.5}) together with Eq.~(\ref{e16.5d})]
and compare the result with the nonperturbative rate $w_1$ calculated
from the exact temporal evolution of $P_B(t)$
[Eq.~(\ref{e16.5c}) together with Eq.~(\ref{e16.5d})].
Noting that \mbox{$P_A(t_0)$ $\!\simeq$
$\!\exp(-\Gamma_{A^\ast A}t_0)$}, we find for the three
cases considered in Eqs.~(\ref{e16.7}) -- (\ref{e16.7a})
the following results.
\begin{itemize}
\item[(i)]
\mbox{$\Gamma_{A^\ast A}$ $\!\gg$ $\!\Gamma_{B^\ast B}$}:
\begin{equation}
\label{e28.4a}
    P_A(t_0) \simeq {\textstyle\frac{1}{4}},
    \quad
    \frac{w_1}{w} \simeq 1.
\end{equation}
\item[(ii)]
\mbox{$\Gamma_{A^\ast A}$ $\!=$ $\!\Gamma_{B^\ast B}$}:
\begin{equation}
\label{e28.4b}
    P_A(t_0) \simeq e^{-\left(2-\sqrt{2}\right)},
    \quad
    \frac{w_1}{w} \simeq \sqrt{2}-1 \simeq 0.41.
\end{equation}
\item[(iii)]
\mbox{$\Gamma_{B^\ast B}$ $\!\gg$ $\!\Gamma_{A^\ast A}$}:
\begin{equation}
\label{e284c}
    P_A(t_0) \simeq 1,
    \quad
    \frac{w_1}{w} \simeq {\textstyle\frac{1}{4}}\,.
\end{equation}
\end{itemize}

Agreement between $w_1$ and $w$ is observed in the first
case, where $\Gamma_{B^\ast B}$ is sufficiently small.
With increasing value of $\Gamma_{B^\ast B}$ an increasing
discrepancy between $w_1$ and $w$ is observed.
The reason can be seen in the fact that there are two processes
which simultaneously drive atom $B$: the energy transfer
from $A$ to $B$ and the spontaneous decay. Hence, $w_1$
is actually the rate of both processes combined,
not the bare rate of energy transfer. This explains why
$w_1$ is typically smaller than $w$ and why the discrepancy between
them becomes more substantial for enhanced
spontaneous decay of atom $B$.
To roughly compensate for the first effect,
the ratio $w_1/w$ may be multiplied by
$\exp(\Gamma_{B^\ast B}t_0)$. The values of the
so corrected ratio are then $\simeq 1$ for the cases
(i) and (iii), and $\simeq 0.74$ for the case (ii).


\subsection{Strong atom-field coupling}
\label{strongcoupling}

For the sake of transparency, we again consider identical atoms
that have equivalent positions and dipole orientations
with respect to the material surroundings
so that Eqs. (\ref{e16.4a}) and (\ref{e16.4b}) hold.
Introducing the probability amplitudes
\begin{equation}
\label{e20}
      C_\pm (t) = 2^{-\frac{1}{2}}
      \left[ C_A(t) \pm C_B(t) \right]
      e^{\mp i\delta_{A^\ast B}t}
\end{equation}
of the superposition states
$|\pm\rangle$ $\!=$ $\!2^{-1/2} \left( |U_A\rangle\pm|U_B\rangle \right)$,
from Eqs.~(\ref{e9}) we find that the equations for $C_+(t)$
and $C_-(t)$ decouple,
\begin{equation}
\label{e20.1}
        \dot{C}_{\pm}(t) =
        \int_0^t {\rm d}t' \Ints \!{\rm d}\omega \,
        K_\pm(t,t';\omega)\,e^{\mp i\delta_{A^\ast B}(t-t')}\, C_\pm(t'),
\end{equation}
where
\begin{equation}
\label{e20.2}
       K_\pm(t,t';\omega) = K_{A^\ast A}(t,t';\omega)
       \pm K_{A^\ast B}(t,t';\omega) .
\end{equation}

Let us restrict our attention to the case when
the absolute value of the two-atom term $K_{A^\ast B}(t,t';\omega)$
is of the same order of magnitude as the absolute value of the
single-atom term $K_{A^\ast A}(t,t';\omega)$, so that there is
a strong contrast in the magnitude of $K_+(t,t';\omega)$
and $K_-(t,t';\omega)$. Assuming that the
imaginary part of the Green tensor in the resonance
region of strong RAFI
has a Lorentzian shape, with $\omega_m$
and $\Delta\omega_m$ being  the central frequency and
the half width at half maximum  respectively, we can
perform the frequency integral in Eq.~(\ref{e20.1}) in
a closed form in a similar way as in the single-atom case
\cite{Scheel99,Ho00}.
In particular for exact resonance, i.e.,
\mbox{$\omega_m$ $\!=$ $\!\tilde{\omega}_A\mp\delta_{A^\ast B}$},
we derive
\begin{equation}
\label{e21}
       C_\pm(t) = 2^{-\frac{1}{2}}
       e^{-\Delta\omega_m t/2} \cos\!\left(\Omega_\pm t/2\right)
\end{equation}
($\Omega_\pm$ $\!\gg$ $\!\Delta\omega_m$),
where
\begin{equation}
\label{e20.4}
      \Omega_\pm = \sqrt{2\Gamma_\pm\Delta\omega_m}\,,
      \quad \Gamma_\pm = \Gamma_{A^\ast A}\pm\Gamma_{A^\ast B}
\end{equation}
with $\Gamma_{A^\ast A}$ and $\Gamma_{A^\ast B}$ being defined according
to Eq.~(\ref{e16}) with $\omega_m$ in place of
\mbox{$\tilde{\omega}_A$ $\!(\simeq$ $\!\tilde{\omega}_B)$}.
For the probability amplitudes
of the remaining states $|\mp\rangle$, which are weakly coupled to the field,
we obtain
\begin{equation}
\label{e22}
       C_\mp(t) = 2^{-\frac{1}{2}} e^{-\Gamma_\mp t/2}
\end{equation}
($\Omega_\mp$ $\!\ll$ $\!\Delta\omega_m$).
It then follows that
\begin{eqnarray}
\lefteqn{
       P_{A(B)}(t)=\textstyle\frac{1}{ 4} \Bigl[
       e^{-\Gamma_\mp t} + e^{-\Delta\omega_m t}
       \cos^2\!\left(\Omega_\pm t/2\right)
}
\nonumber\\[1ex]&&\hspace{-3ex}
\label{e22.1}
     +(-) \,2e^{-(\Delta\omega_m  +\Gamma_\mp)t/2}
     \cos\!\left(\Omega_\pm t/2\right)
     \cos\!\left(2\delta_{A^\ast B}t\right) \Bigr].
\end{eqnarray}
Note that in the equations given above the upper (lower) signs
refer to the case where the state $|+\rangle$ ($|-\rangle$) is
strongly coupled to the medium-assisted field.

For small damping, between three typical cases
can be distinguished.
\begin{itemize}
\item[(i)] $4|\delta_{A^\ast B}|$ $\!\gg$ $\!\Omega_\pm$,
$t$ $\!\ll$ $\!2/\Omega_\pm$:
\begin{eqnarray}
\label{e23}
      P_A(t) &=& \cos^2(\delta_{A^\ast B}t),
\\
\label{e23.a}
      P_B(t) &=& \sin^2(\delta_{A^\ast B}t).
\end{eqnarray}
\item[(ii)] $4|\delta_{A^\ast B}|$ $\!\simeq$ $\!\Omega_\pm$,
$t$ $\!\ll$ $\!1/|2\delta_{A^\ast B}$ $\!-$ $\!\Omega_\pm/2|$:
\begin{eqnarray}
\label{e24}
      P_A(t) &=& \textstyle\frac{1}{ 4}
      \left[1+3\cos^2\!\left(\Omega_\pm t/2\right)\right],
\\
\label{e24.a}
      P_B(t) &=& \textstyle\frac{1}{ 4}
      \left[1-\cos^2\!\left(\Omega_\pm t/2\right)\right].
\end{eqnarray}
\item[(iii)] $4|\delta_{A^\ast B}|$ $\!\ll$ $\!\Omega_\pm$,
$t$ $\!\ll$ $\!1/|2\delta_{A^\ast B}|$:
\begin{eqnarray}
\label{e25}
      P_A(t) &=& \cos^4\!\left(\Omega_\pm t / 4\right),
\\
\label{e25.a}
      P_B(t) &=& \sin^4\!\left(\Omega_\pm t / 4\right).
\end{eqnarray}
\end{itemize}
{F}rom Eqs.~(\ref{e23}) -- (\ref{e25.a}) the following
time-averaged probabilities $\bar{P}_A$, $\bar{P}_B$,
and \mbox{$\bar{P}_L$ $\!=$ $\!1$ $\!-$ $\!\bar{P}_A$ $\!-$
$\!\bar{P}_B$} are obtained, with the respective time
integral being taken over one cycle.
(i) \mbox{$\bar{P}_A$ $\!=$ $\!\bar{P}_B$ $\!=$ $\frac{1}{ 2}$}
and \mbox{$\bar{P}_L$ $\!=$ $\!0$}. The excitation energy is
periodically exchanged between the two atoms through virtual
field excitations exclusively.
(ii) \mbox{$\bar{P}_A$ $\!=$ $\!\frac{5}{8}$}, \mbox{$\bar{P}_B$
$\!=$ $\frac{1}{8}$}, and \mbox{$\bar{P}_L$ $\!=$
$\!\frac{2}{8}$}.
The two exchange channels -- one channel through virtual and the other
one through real field excitations -- compete with each other and
destructively interfere, leading to a partial trapping of
the excitation energy in atom $A$.
(iii) \mbox{$\bar{P}_A$ $\!=$ $\!\bar{P}_B$ $\!=$ $\frac{3}{8}$},
\mbox{$\bar{P}_L$ $\!=$ $\!\frac{2}{8}$}.
The interatom energy exchange is dominantly mediated by real
field excitations.


\section{Power spectrum}
\label{spectrum}

Let us finally address the problem of the influence of the
RDDI on the power spectrum of the light emitted
during the decay-assisted energy exchange.
To calculate the (physical) spectrum,
we apply the formula (see, e.g., \cite{Vogel01})
\begin{eqnarray}
\label{e28.5}
\lefteqn{
          S({\bf r},\omega_{\rm S},T) =
          \int_0^T {\rm d}t_2\int_0^T {\rm d}t_1
	  \Bigl[e^{-i\omega_{\rm S}(t_2-t_1)}
}
\nonumber\\[1ex]&&\hspace{17ex}\times\,
          \left\langle \hat{\bf E}^{(-)}({\bf r},t_2)
          \hat{\bf E}^{(+)}({\bf r},t_1) \right\rangle\Bigr],
\end{eqnarray}
where $\omega_{\rm S}$ is the setting frequency of the (ideal)
spectral apparatus, $T$ is the operating-time interval of the detector,
and
\begin{equation}
\label{e28.6}
\hat{\bf E}^{(+)}({\bf r})
   = \int_{0}^{\infty} {\rm d}\omega\,
   \underline{\hat{\bf E}}({\bf r},\omega),
\end{equation}
\begin{equation}
\label{e28.6a}
\hat{\bf E}^{(-)}({\bf r})
   = \left[\hat{\bf E}^{(+)}({\bf r})\right]^\dagger.
\end{equation}


\subsection{Weak atom-field coupling}
\label{spectrum_wc}

For weak RAFI, we derive, on basing on
Eqs.~(\ref{e16.1}), (\ref{e16.2}), (\ref{e16.4a}), and (\ref{e16.4b}),
\begin{eqnarray}
\label{e18.1}
\lefteqn{
     S({\bf r},\omega_S,T \rightarrow \infty) = \frac{1}{ 4}
     \biggl|
     \frac{{\bf F}_A+{\bf F}_B }{ \Delta\omega_S
     + \delta_{A^\ast B} +i\Gamma_+/2}
}\qquad
\nonumber\\[1ex]&&\hspace{10ex}
     +\, \frac{{\bf F}_A-{\bf F}_B}{ \Delta\omega_S
     - \delta_{A^\ast B} +i\Gamma_-/2}
     \biggr|^2,
\end{eqnarray}
where
\begin{equation}
\label{e18.2}
           \Delta\omega_S=\omega_S-\tilde{\omega}_A
\end{equation}
and \mbox{($A'$ $\!=$ $\!A,B$)}
\begin{equation}
\label{e18.4}
          {\bf F}_{A'} =
          \frac{\tilde{\omega}_{A'}^2 }{ \pi\epsilon_0 c^2}
          \Ints {\rm d}\omega\,{\rm Im}\,\bm{G}({\bf r},{\bf r}_{\rm A'},
          \omega)\,{\bf  d}_{A'}\zeta(\tilde{\omega}_{A'}-\omega).
\end{equation}
%
\begin{figure}[htb]
\includegraphics[width=1\linewidth]{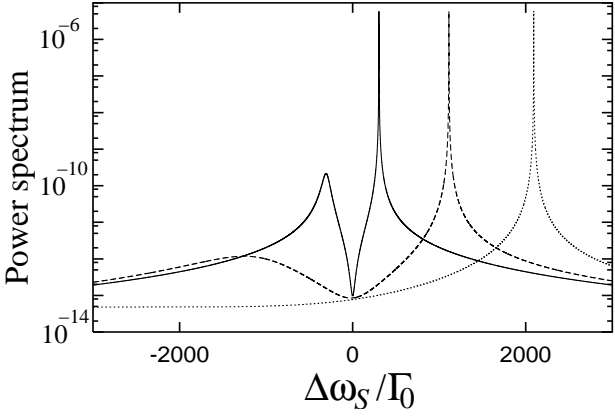}
\caption{
\label{specw}
The power spectrum
$S[64\pi\varepsilon_0 c^3\Gamma_0/(3 \hbar\tilde{\omega}_A^3)]$
in the weak coupling regime for two atoms situated near
a dielectric microsphere of single-resonance Drude-Lorentz-type
[$\omega_{\rm T}$, transverse frequency;
\mbox{$\omega_{\rm P}$ $\!=$ $\!0.5\,\omega_{\rm T}$},
plasma frequency;
$\gamma$ $\!=$ $\!10^{-6}\,\omega_{\rm T}$, absorption parameter;
$d$ $\!=$ $\!20\,\lambda_{\rm T}$, sphere diameter ($\lambda_{\rm T}$
$\!=$ $\!2\pi c/\omega_{\rm T}$);
$\omega_m$ $\!=$ $\!1.0504867\,\omega_{\rm T}$;
$\Delta\omega_m$ $\!\simeq$ $\!5$ $\!\times$ $\!10^{-7}\,\omega_{\rm T}$;
$\Delta r_A$ $\!\equiv$ $\!r_A$ $\!-$ $\!0.5\,d$
$\!=$ $\!\Delta r_B$ $\!=$ $\!0.02\,\lambda_{\rm T}$,
distance of the atoms from the sphere surface;
${\bf d}_A$ $\!=$ $\!{\bf d}_B$, radially oriented
real transition dipole moments;
\mbox{$R$ $\!=$ $\!20.04\,\lambda_{\rm T}$}, interatomic distance;
the detector is positioned at $r$ $\!=$ $\!20\lambda_{\rm T}$
on the radial line connecting the sphere
center and atom $A$;
$\tilde{\omega}_A$ $\!=$ $\!1.05048\,\omega_{\rm T}$ (solid line),
$\tilde{\omega}_A$ $\!=$ $\!1.050485\,\omega_{\rm T}$ (dashed line), and
$\tilde{\omega}_A$ $\!=$ $\!1.05048621\,\omega_{\rm T}$ (dotted line).
}
\end{figure}
%
Note that replacing \mbox{$\zeta(\tilde{\omega}_{A'}$
$\!-$ $\!\omega)$} with \mbox{$\pi\delta(\tilde{\omega}_{A'}$ $\!-$
$\!\omega)$} would be too rough here.
Equation (\ref{e18.1}) reveals that the
RDDI may result in a doublet structure (two
asymmetric lines at \mbox{$\tilde{\omega}_A$
$\!\mp$ $\!\delta_{A^\ast B}$}, with $\Gamma_\pm$ and
\mbox{$|{\bf F}_A$ $\!\pm$ $\!{\bf F}_B|^2$} being the widths and weights
respectively). The line separation is seen to be twice the
RDDI strength. A system of two two-level atoms, one of them
initially excited, is obviously equivalent to a three-level system
with two upper dressed states $|\pm\rangle$.
The doublet structure can be understood
as a result of the transitions of the dressed states to the ground state.

As can be seen from Eq.~(\ref{e18.1}), an experimental observation
of the doublet structure of the emitted light requires
a delicate balancing act. The interatomic distance should not be too large
to provide a reasonable level splitting, but it should not be too
small to avoid \mbox{$|{\bf F}_A|$ $\!=$ $\!|{\bf F}_B|$}, i.e.,
quenching of one of the two lines. It is worth noting that
the presence of macroscopic bodies may facilitate the detection
of the doublet, because it offers the possibility of realizing
strong RDDI even for interatomic
distances much larger than the wavelength.

Another interesting feature
is that, according to Eq.~(\ref{e20.4}), either $\Gamma_+$ or
$\Gamma_-$ can be much smaller than \mbox{$\Gamma_{A^\ast A}$
$\!(=$ $\!\Gamma_{B^\ast B}$)}. That is, the resonant dipole-dipole
interaction can give rise to an ultranarrow spectral line, albeit
each time at the expense of the other line of the pair.
If, e.g., a single atom is placed sufficiently near a microsphere,
its spontaneous decay may be suppressed, with the emission line
being accordingly narrowed. Compared to the emission line of
a single atom, one line of the doublet observed
for two atoms being present may be further narrowed by
several orders of magnitude \cite{Ho01a}.

The RDDI-induced asymmetric splitting of the power spectrum
is illustrated in Fig.~\ref{specw} for
two atoms near a microsphere.
The solid line in Fig.~\ref{specw} corresponds to an
off-resonant atomic transition
with not too strong
RDDI and moderate
contrast between $\Gamma_+$ and $\Gamma_-$. As
the atomic transition frequency
$\tilde{\omega}_A$ approaches (from the off-resonance side)
the frequency, where the absolute value of the RDDI strength
reaches its maximum, the two spectral lines move
apart from each other, with one of them gradually getting
so much broaden that it eventually disappears (dashed and dotted
lines). As
$\tilde{\omega}_A$ shifts further towards the resonance line center,
the RDDI weakens while
the contrast between $\Gamma_+$ and $\Gamma_-$ increases even more.
As a result, the singlet structure persists with the peak
moving back to the central position (not shown).


\subsection{Strong atom-field coupling}
\label{spectrum_sc}

\begin{figure}[thb]
\includegraphics[width=1\linewidth]{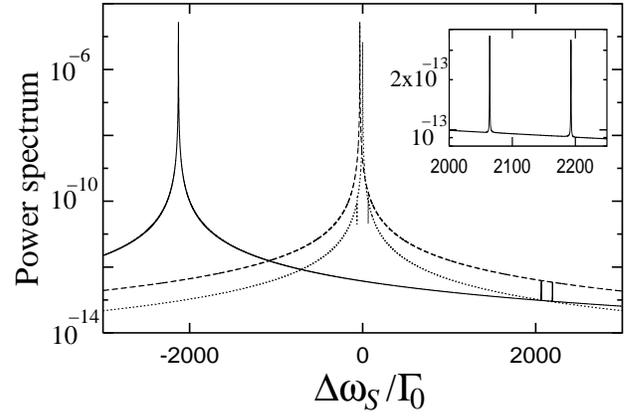}
\caption{
\label{specstr}
The power spectrum
$S[64\pi\varepsilon_0 c^3\Gamma_0/(3 \hbar\tilde{\omega}_A^3)]$
in the strong coupling regime for the three sets of parameters
[$\Gamma_0$ $\!=$ $\!10^{-6}\,\omega_{\rm T}$,
$\Omega_+$ $\!\simeq$ $\!128\,\Gamma_0$;
(i) $\tilde{\omega}_A$ $\!=$ $\!1.04835747\,\omega_{\rm T}$,
$R$ $\!\simeq$ $\!0.01\,\lambda_{\rm T}$,
$\delta_{AB}$ $\!\simeq$ $\!-$ $\!2129\,\Gamma_0$ (solid line);
(ii) $\tilde{\omega}_A$ $\!=$ $\!1.05045444\,\omega_{\rm T}$,
$R$ $\!\simeq$ $\!0.027\,\lambda_{\rm T}$,
$\delta_{AB}$ $\!\simeq$ $\!-$ $\!32.2\,\Gamma_0$ (dashed line);
(iii) $\tilde{\omega}_A$ $\!=$ $\!\omega_m$ $\!=$
$\!1.0504867\,\omega_{\rm T}$,
$R$ $\!=$ $\!20.04\,\lambda_{\rm T}$,
$\delta_{AB}\simeq 0$] (dotted line). Other parameters are the
same as in Fig. \ref{specw}.
}
\end{figure}
%
For strong RAFI, Eq.~(\ref{e18.1}) changes to,
on basing on Eqs.~(\ref{e21}) and (\ref{e22}),
\begin{eqnarray}
\label{e25.1}
\lefteqn{
     S({\bf r},\omega_S,T \rightarrow \infty)
     = \frac{1}{ 4}\biggl| ({\bf W}_A\pm {\bf W}_B)
}
\nonumber\\[1ex]&&\hspace{2ex}\,\times
     \biggl(
     \frac{1}{  \Delta\omega_S\pm\delta_{A^\ast B}
      +\Omega_\pm/2+i\Delta\omega_m/2}
\nonumber\\[1ex]&&\hspace{4ex}
     -\,\frac{1}{  \Delta\omega_S\pm\delta_{A^\ast B}
      -\Omega_\pm/2+i\Delta\omega_m/2}
     \biggr)
\nonumber\\[1ex]&&\hspace{8ex}
     +\,\frac{i({\bf F}_A\mp {\bf F}_B) }{  \Delta\omega_S\mp \delta_{A^\ast B}
      +i\Gamma_\mp/2} \biggr|^2,
\end{eqnarray}
where \mbox{($A'$ $\!=$ $\!A,B$)}
\begin{equation}
\label{e25.2}
      {\bf W}_{A'}=
          \frac{\omega_m^2\Delta\omega_m  }{
          \epsilon_0 c^2 \Omega_\pm}\,
          {\rm Im}\,\bm{G}({\bf r},{\bf r}_{\rm A'},\omega_m)
          \,{\bf d}_{A'}.
\end{equation}
Here the upper (lower) signs again refer to the case where
the state $|+\rangle$ ($|-\rangle$) is strongly coupled to the
medium-assisted field. Eq.~(\ref{e25.1}) reveals that
due to the strong RAFI a doublet observed for
weak atom-field coupling may become a triplet,
with one of the lines of the doublet being split into
two lines. These lines separated by $\Omega_\pm$
have equal widths (which are solely determined by the width of
the medium-assisted field resonance) and equal
weights. Note that their width and weight are different from those
of the third line, which is closely related to a line of
the doublet observed for weak RAFI.
{F}rom Eq.~(\ref{e25.1}) it is also seen that, depending on the
point of observation,
this line or the strong-coupling-assisted doublet can be suppressed
due to the interference effects.

The effect of strong RAFI on the spectrum of the emitted
light is illustrated in Fig.~\ref{specstr} for two atoms near
a microsphere.
Three typical situations are shown, namely (i) strong,
(ii) moderate, and (iii) weak RDDI
(cf. Subsection~\ref{strongcoupling}).
Note that the triplet structure is most pronounced in the
first case.


\section{Concluding remarks}
\label{remarks}

We have developed a fully quantum mechanical theory
to describe both RDDI and RAFI in the presence of arbitrarily
configured dispersing and absorbing material bodies.
In particular, the concept of RDDI works well for arbitrary
atom-field coupling strengths. Though both the single-atom
level shifts and the RDDI strengths result from the off-resonant
atom-field interaction and are formally given by
the similar-looking Eqs. (\ref{e10.1}) and (\ref{e10.2}),
respectively, Eq.~(\ref{e10.1}) applies only to the 
level shifts caused by the scattering part of the Green tensor.
The level shifts that are related to the vacuum field fluctuations
have to be dealt with separately due to the divergence of the
real part of the vacuum Green tensor at equal positions.
In our calculations, they have been thought of as being already
included in the bare atomic transition frequencies.
By contrast, the real part of the Green tensor at different
positions is finite and Eq.~(\ref{e10.2}) can thus be used
to calculate the total RDDI strengths
caused by both vacuum and scattered-field fluctuations.
In the same context, it might be incorrect to apply the
rotating-wave approximation in the study of the RDDI.
While the (scattering-assisted) level shifts
$\delta^+_{A^\ast A}$, which result from 
counterrotating terms, are small in general [cf. Eqs. (\ref{e3.8})
and (\ref{e10.1})], the corresponding RDDI strengths 
$\delta^+_{A^\ast A'(A\neq A')}$  can be significant,
especially when the mutual distances of the dipoles are
very small [cf. Eq.~(\ref{e10.2})].

To demonstrate the usefulness of the theory, we have applied it to
the problem of resonant energy exchange between two atoms.
In particular, we have shown that the RDDI can give rise
to spectral line splittings of the light emitted during the
decay-assisted energy exchange process.
With a locally excited dipole as a probe, the effect may
be employed, e.g., in imaging single atoms on a surface
by scanning near-field optical microscopy techniques.
The (numerical) calculations performed for atoms near
a microsphere can of course be extended to other
geometries of the material surroundings such as
planarly stratified or cylindrical objects. It should also
be mentioned that the theory is also readily
extendable to the case of externally driven atoms.

\vspace*{0.2cm}
\begin{acknowledgments}
This work was supported by the Deutsche Forschungsgemeinschaft.
\end{acknowledgments}



\begin{references}

\bibitem{Ho02}
Ho Trung Dung, L. Kn\"oll, and D.-G. Welsch,
Phys. Rev. A {\bf 65}, 043813 (2002).

\bibitem{Lehmberg70}
R. H. Lehmberg, Phys. Rev. A {\bf 2}, 889 (1970).

\bibitem{Agarwal74}
G. S. Agarwal, {\it Quantum Optics}, Vol. 70 of
{\it Springer Tracts in Modern Physics} (Springer, Berlin, 1974).

\bibitem{Craig84}
D. P. Craig and T. Thirunamachandran,
{\it Molecular Quantum Electrodynamics}
(Academic, New York, 1984).

\bibitem{Sorensen02}
A. S. S{\o}rensen and K. M{\o}lmer, quant-ph/0202073.

\bibitem{Barenco95}
A. Barenco, D. Deutsch, A. Ekert, and R. Jozsa, Phys. Rev. Lett.
{\bf 74}, 4083 (1995).

\bibitem{Lukin00}
M. D. Lukin and P. R. Hemmer, Phys. Rev. Lett. {\bf 84}, 2818 (2000).

\bibitem{Kurizki96}
G. Kurizki, A. G. Kofman, and V. Yudson, Phys. Rev. A {\bf 53}, R35 (1996).

\bibitem{Knoll01}
L. Kn\"{o}ll, S. Scheel, and D.-G. Welsch,
in {\it Coherence and Statistics of Photons and Atoms},
edited by J. Pe\v{r}ina (John Wiley \& Son, New York, 2001), p. 1.

\bibitem{Juzeliunas94}
G. Juzeli\=unas and D. L. Andrews, Phys. Rev. B {\bf 50}, 13371 (1994).

\bibitem{Ho01a}
Ho Trung Dung, S. Scheel, L. Kn\"oll, and D.-G. Welsch,
J. Opt. B: Quant. Semiclass. Opt. {\bf 4}, S169 (2002).

\bibitem{Forster48}
T. F\"orster, Ann. Phys. (Leipzig) {\bf 1}, 55 (1948).

\bibitem{Dexter53}
D. L. Dexter, J. Chem. Phys. {\bf 21}, 836 (1953).

\bibitem{Scheel99}
S. Scheel, L. Kn\"{o}ll, and D.-G. Welsch,
Phys. Rev. A {\bf 60}, 4094 (1999);

\bibitem{Ho00}
Ho Trung Dung, L. Kn\"oll, and D.-G. Welsch,
Phys. Rev. A {\bf 62}, 053804 (2000).

\bibitem{Vogel01}
W. Vogel, D.-G. Welsch, and S. Wallentowitz,
{\it Quantum Optics, An Introduction}
(Wiley-VCH, Berlin, 2001).

\end{references}
\end{document}